# Water vapor assisted sintering of silver nanoparticle inks for printed electronics

Justin. Bourassa, Alex Ramm, James Q. Feng, and Michael J. Renn

Optomec, Inc., 2575 University Avenue, #135, St. Paul, MN 55114, USA

**Email:** jfeng@optomec.com

**Abstract**

In printed electronics, conductive traces are often produced by printing inks of silver nanoparticles dispersed in solvents. A sintering process is usually needed to make the printed inks conductive by removing the organic dispersants and allowing metal-to-metal contacts among nanoparticles for atomic diffusion and neck formation. It has been challenging to sinter silver nanoparticle inks in a thermal oven at a temperature < 150$^o$ C to avoid thermal damage to the plastic substrate while achieving desired conductivity. This work presents a simple yet effective way to sinter a silver nanoparticle ink below 120$^o$ C (even at 80$^o$ C) by exposing the printed ink to water vapor in the oven. The results consistently show a significant reduction of line resistivity for the samples sintered in a moist oven compared to those sintered in a dry oven. Hence, solvent vapor-assisted sintering of metal nanoparticle inks can become an enabling approach to broaden product range of printed electronics.

**Introduction**

Functional inks consisting of metal nanoparticles dispersed in solvents are commonly used in the printed electronics industry to fabricate conductive patterns on plastic substrates [1]. To prevent agglomeration of those nanoparticles, metal particle binding thiols and polymers (also called the "capping agent") are used as dispersants for stabilizing the dispersion [2, 3]. Such metal nanoparticle inks can be deposited onto substrates according to designed circuit pattern by a number of printing techniques like screen printing, syringe-needle type dispenser, inkjet, Aerosol Jet®, to mention a few. After printing and solvent evaporation, the metal nanoparticle materials usually do not become adequately conductive. The particle-ligand bonds must be broken for removal of polymeric ligands between contacting particle surfaces to establish metal-to-metal atomic diffusion among particles, because the movement of electrons between metal particles can be prevented by an organic layer as thin as a few nanometers [4].

As a common practice, sintering the printed metal nanoparticle ink at an elevated temperature (e.g., in a thermal oven) can make it electrically conductive, while the polymeric ligands might be removed by thermal decomposition (known as pyrolysis) and vaporization. Simultaneously, the contacting metal nanoparticles are fused together via interfacial atomic diffusion and neck formation [5]. When heated to 70° C the particles in a silver nanoparticle ink appeared to touch each other as the organic molecules started moving away from particle surface; but the electrical resistivity reached minimum value only at a temperature > 200° C and then increased with further heating to temperature beyond 250° C due to growing pore size with reduced connectivity among coalesced large metal domains [6].

Sintering silver nanoparticle inks in a thermal oven, although conceptually straightforward, poses practical limitations for printed electronics on common low-cost polymer substrates (e.g., polycarbonate and polyethylene terephthalate) with glass transition temperature well below 200° C. To avoid thermal damage to the plastic substrate, the printed parts should not be sintered above the heat deflection temperature of substrate. On the other hand, it is often desired to obtain lowest possible resistivity of the sintered ink material for the desired electronic performance. When producing functional electronic devices, it is also important to have adequate adhesion and cohesion of the sintered nanoparticle material on a substrate, which may also require increased sintering temperature. Thus, development of metal nanoparticle inks for low-temperature sintering has been an ongoing endeavor in the printed electronics industry. By incremental improvements of ink formulations, some modern silver nanoparticle inks can now be sintered at temperatures as low as about 100° C, with an impractically long duration (e.g., days) just to achieve marginal values of electrical conductivity.

Despite years of intensified research efforts, the ideal metal nanoparticle ink and its sintering process still remain elusive. Besides simply heating the printed parts in a thermal oven, other sintering methods have also been investigated by using localized laser beam [8 – 10], pulsed light [11], microwave [12], electrical current [13], and so on. Some recent findings have also suggested possibilities for sintering silver nanoparticles even at room temperature when utilizing oppositely charged polyelectrolytes [14], or solvent absorbing coating on substrate (which also indicated the beneficial effect of humidity) [15], or destabilizing agent activated during solvent removal [16], or potassium chloride solution [17], etc. However, the method of thermal oven sintering is still widely used in printed electronics production for operational simplicity and convenience.

In this work, we present a process of thermal oven sintering by simply placing a measured amount of water in a bottom tray when putting in the samples. The presence of water vapor while sintering has been found to significantly reduce electrical resistivity of a commercially available silver nanoparticle ink (Sicrys[TM] I60PM-116 by PV Nanocell), when compared to samples sintered without water vapor. When this ink is printed and sintered on polycarbonate substrates, measurements according to ASTM D3359-09 show that all the samples pass 5B adhesion test rating even after exposure to "salt spray" in a foggy chamber of 35° C for 48 hours per ASTM B117-11.

**Experimental methods**

As is typical with printed electronics, the silver nanoparticle ink is deposited according to the test pattern on a substrate by a printing process, and then sintered in a thermal oven with adjustable process settings for temperature, duration, and amount of water. In the present work, an Aerosol Jet® direct-write system[7] is used for printing and a Cole-Parmer® vacuum oven (StableTemp Model 281A) is used for thermal sintering. The printed samples are sintered in an oven chamber with a tray on the bottom containing a measured amount of water, at a set temperature under ambient pressure without active gas pumping.

The test samples for resistivity evaluation are printed with consistent line length $L$, width $W$ and cross-section area $A_x$ with probe pads for four-point resistance measurement (as shown in figure 1), according to a given mass output from the Aerosol Jet® deposition nozzle. After being sintered, the resistance of each printed line $R$ is measured with the four-point probe method. The value of resistivity, $\rho$, is then determined according to

$$\rho = R A / L \quad , \tag{1}$$

where the value of cross-section area, $A$, may be considered simply as that from the measured line profile, $A_x$, in figure 2, which could vary depending upon the sintering condition and packing density of the nanoparticles.

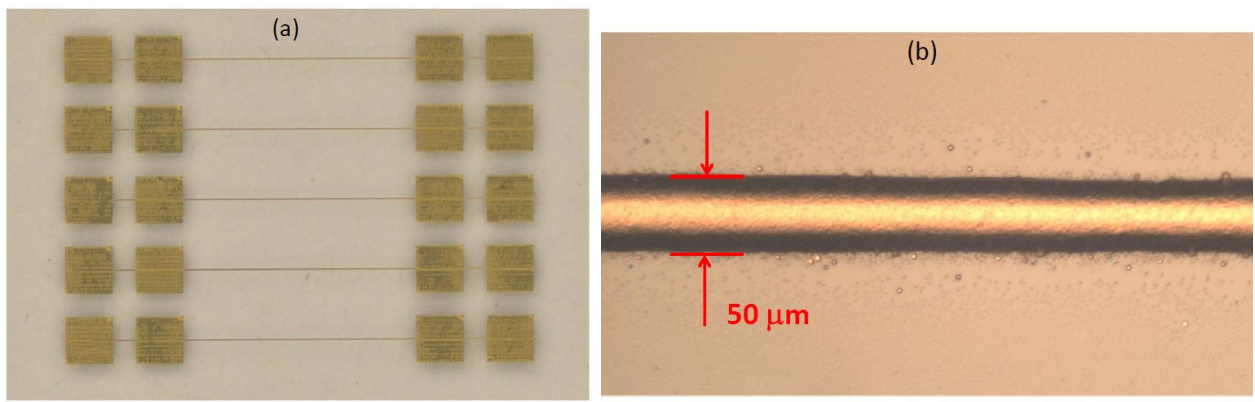

Figure 1. (a) Aerosol Jet® printed lines on glass with probe pads for four-point resistance measurement, and (b) optical microscope image of a line segment. The line width is measured as 50 microns on a micrometer controlled stage.

An alternative way for determining $A$ is to use a calculation based on the metal mass deposition rate, $\mu$, (which can be maintained at a consistent value for hours with the newly developed Aerosol Jet® print engine) and the bulk metal density, $\rho_m$, with the known print speed, $u$, as

$$A_m = \mu / (u\, \rho_m) \quad . \tag{2}$$

The value of $A_m$ from (2) actually corresponds to the effective cross-section area of the metal material, whereas $A_x$ is the geometric cross-section area of the porous material. Thus, using $A_m$ determined by (2) for $A$ in (1) removes the often uncontrollable variable of particle packing density. For the sample shown in figure 2, as printed with a metal mass output rate $\mu = 0.75$ mg/min (or 0.0125 mg/s) at a speed of $u = 10$ mm/s, using $\rho_m = 10.5$ mg/mm$^3$ for bulk silver yields $A_m = 119$ µm$^2$ according to (2), whereas the geometric cross-section area is measured as $A_x = 163$ µm$^2$ (which corresponds to a density of 7.7 mg/mm$^3$ for the sintered nanoparticle material.)

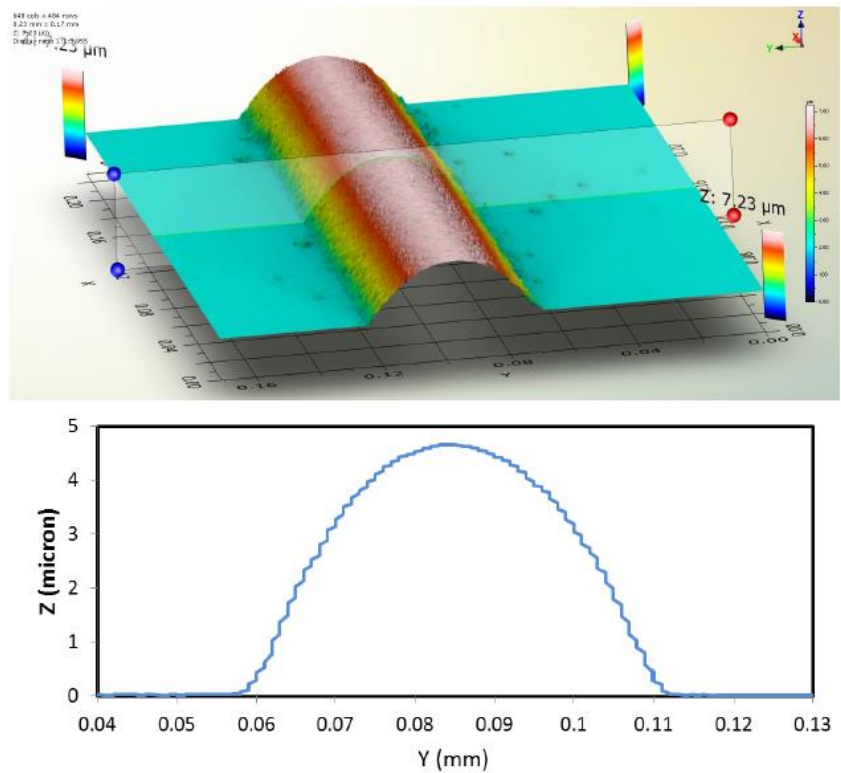

Figure 2. Optical profilometry results of the Aerosol Jet® printed line in Fig. 1 using the Filmetrics Profilm3D based on white light interferometry. It is shown that the line is 0.05 mm (or 50 µm) wide with a cross-section area of $A_x = 163$ µm$^2$ when sintered at 200°C for 60 minutes with 10 grams of water placed in the oven tray.

**Results and discussion**

The effect of having 10 grams of water placed in the bottom tray inside the 18 liters oven chamber is shown in table 1, for oven temperature set at 120°C and 200°C. The values of relative resistivity, in units of the bulk silver value (1.58 µΩ cm), are calculated according to (1) from measured resistance $R$, line length $L$ = 10 mm, and the cross-section area $A = A_m$ determined by (2). After sintered at 120° C with water vapor for an hour, the line resistivity becomes even less than that sintered at 200° C in a dry oven. The relative resistivity values are measured as 2.29 and 5.52 for printed lines sintered at 150° C for 60 minutes with and without water vapor, respectively. Even at 80° C the relative resistivity can become 9.39 if sintered with water vapor for four hours. Clearly, the resistivity values of printed lines sintered with water vapor are significantly reduced from that in dry oven, especially at lower sintering temperature.

Table 1. Measured resistivity values (in units of the bulk silver resistivity) under different sintering conditions.

| T (°C)[a] | 30 (min) | 60 (min) | 120 (min) | Water (gram)[b] |
|---|---|---|---|---|
| 120 | --- | > 720 | 22.62 | 0 |
| 120 | 5.90 | 3.24 | 3.45 | 10 |
| 200 | 4.39 | 4.48 | 4.81 | 0 |
| 200 | 2.05 | 2.13 | 2.07 | 10 |

[a]T is the oven set temperature.
[b]Amount of water placed in the bottom tray in the oven chamber.

To illustrate the effect of amount of water placed in the oven chamber, figure 3 shows the measured resistivity of printed lines sintered at 120° C for one hour as a function of water amount. Interestingly, the result of sintering with this process is rather insensitive to the amount of water placed in the oven tray as long as more than 10 grams of water (in an 18 liter oven chamber) is provided. Thus, a robust sintering process can be easily implemented by following a simple rule of > 0.5 grams of water per liter of oven chamber volume.

The water vapor mass, $M$, in the oven chamber of volume, $V$, may be estimated in terms of the liquid water mass, $M_w$, placed in the tray and the maximum water vapor mass (based on Amagat's law of partial volume) $M_{max} = P V / (R_v T)$ at pressure, $P$, and temperature, $T$, (with $R_v$ = 461 J K$^{-1}$ kg$^{-1}$ denoting the gas constant for water vapor) as

$$M = M_{max} [1 - (1 - M_i / M_{max}) / \exp(M_w / M_{max})] \quad , \tag{3}$$

where $M_i$ is the initial water vapor mass inside the oven chamber (which may often be assumed as zero). Here we have assumed the partial volume of vaporized liquid water to displace the same volume of gas containing a mass density of water vapor $M / V$, such that $P$ is maintain at a constant value. With $P = 10^5$ Pa, $V = 0.018$ m$^3$ and $T = 393$ K (for 120° C), we have $M_{max}$ ~ 0.01 kg (or 10 grams of liquid water). If we place $M_w$ = 10 grams of water in the tray, the oven chamber would contain $M$ = 6.3 grams of water vapor according to (3) for $M_i$ = 0. The value of $M$ will approach $M_{max}$ (= 10 grams) when $M_w$ / $M_{max}$ >> 1.

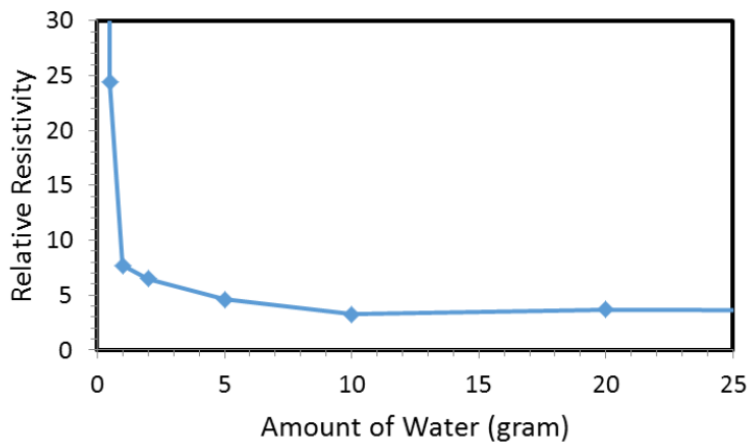

Figure 3. Measured values of line resistivity (in units of the bulk silver value) as a function of amount of water placed in the oven chamber, for an hour sintering at $T$ = 120° C. The flat part for constant resistivity actually extends up to 50 grams of water (as tested), though certain amount of liquid water may remain in the tray at end of sintering when more than 10 grams of water is used.

The effect of water vapor on the structure of sintered nanoparticles can also be revealed from the scanning electron microscopy (SEM) images in figure 4, showing the presence of water vapor enhances nanoparticle coalescence to form much larger connected (bright) regions of metallic silver. Since the temperature is only 120° C, thermal decomposition and vaporization of the organic capping agent are not expected to occur, as evidenced by organic materials in the dark regions.

When describing the basic mechanisms for sintering the metal nanoparticle inks in printed electronics at a temperature (e.g., ~150° C) much lower than the bulk metal melting point $T_m$ (e.g., 960° C for silver), most researchers would point to the phenomenon of "melting point depression". However, quantitative results show that significant reduction of melting temperature only occurs for nanoparticles of diameters below 10 nm, while the size of silver nanoparticles in the PV Nanocell ink used here is typically greater than 50 nm with estimated melting point still above 900° C [18]. In the powder metallurgy process (with larger particles though), the sintering temperature for a single-component powder is usually higher than $T_m$ / 2. Then it is difficult to expect sintering of silver nanoparticle inks at $T$ ~ 150° C (as often observed with the silver nanoparticle inks for printed electronics[5]), if based on the mechanism of melting point depression alone.

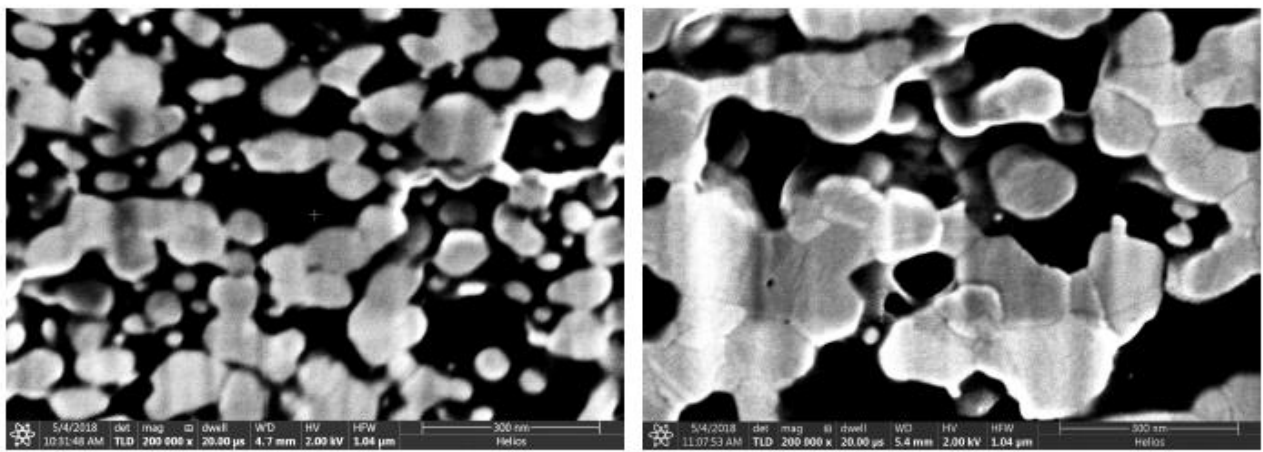

Figure 4. The SEM images of sintered silver nanoparticles at 120°C for two hours in a dry oven (left) and that with water vapor (right). The bright regions are the metallic silver while the dark regions represent the nonconductive material such as organic capping agent and voids.

As their size decreases, nanoparticles will have increased surface-to-volume ratio (which is inversely proportional to the particle diameter). Therefore, there are relatively more atoms on the surface of the nanoparticle. Those surface atoms are generally more energetic, due to weaker intermolecular bonds than those interior ones. Liquefaction of solid surface below $T_m$ (or "surface pre-melting") had indeed been directly observed [19]. Based on several recent publications [14 – 17], the phenomenological knowledge seems to suggest that the key for successful sintering of metal nanoparticle inks at low temperature is to facilitate the removal of the organic capping agent from the contacting gap between adjacent particles. Once the metal particle-to-particle contact is established, the intermolecular attraction due to Van der Waals forces may be sufficient to bring about effective coalescence of particles. According to a computational model with Lennard-Jones force law, contacting spherical particles under the influence of attractive intermolecular forces would deform to have finite (nonzero) contact area [20, 21]. Across such a contact area, atomic diffusion is expected to happen spontaneously for neck growth between coalescing particles even at room temperature, as revealed by molecular dynamics simulations [22]. Hence, finding the most convenient way to remove the organic capping agent from the particle contacting gap can be an effective approach to enable low-temperature sintering of metal nanoparticle inks.

Generally speaking, the primary outcome of sintering is a reduction of surface area due to atomic motion for neck growth and formation of interparticle bonds. When a liquid phase coexists with a particulate solid at the sintering temperature, the rate of interparticle bonding is usually increased by liquid capillary action which has an equivalent effect of using external pressure as with hot pressing (to densify powder compacts while lowering $T_m$). In fact, many commercial ceramic products are actually fabricated with a liquid phase present during sintering [23]. Thus, it may be understandable to find enhanced sintering outcomes with certain types of silver nanoparticle inks by introducing water vapor in the oven chamber. With the small pore sizes among nanoparticles, capillary condensation is expected to happen even at low vapor pressure in the diminishing gaps between contacting particles and the capillary attraction due to a wetting liquid to expedite neck growth during sintering process.

**Concluding remarks**

In summary, we have demonstrated an easy and robust method for successful sintering a silver nanoparticle ink in a thermal oven at temperature below $120^{\circ}$ C (event at $80^{\circ}$ C) by simply placing an adequate amount of water therein. The obtained values of electrical resistivity of sintered ink with water vapor assistance can become about 3 times the bulk silver resistivity, while maintaining 5B adhesion test rating on polycarbonate substrates even after exposure to "salt spray" by measurements according to ASTM D3359-09. Although results only for the silver nanoparticle ink from PV Nanocell are presented here, we also obtained similar beneficial effects of water vapor-assisted sintering with several other silver nanoparticle inks (e.g., that of Advanced Nano Products, PARU, NovaCentrix, etc.) Using water as the vapor solvent in the present work is mainly for its simplicity in laboratory implementation; other solvents may also be considered depending on particular ink formulations as yet to be explored. Our results suggest a new research direction for solvent vapor-assisted sintering of metal nanoparticle inks at lower temperatures, to enable broadening the product range of printed electronics.

**Acknowledgments**

The authors are grateful for convenient access to the Characterization Facility of University of Minnesota that enabled us to obtain the SEM images of sintered nanoparticles. The effort of Kelley McDonald in producing the SEM images and helpful discussion with Dr. Kurt Christenson are greatly appreciated.

**ORCID iDs**

James Q. Feng  https://orcid.org/0000-0003-4041-3179